\documentclass[prd,floats,aps,twocolumn,epsf,graphicx]{revtex4}
\usepackage{epsfig}
\usepackage{graphicx}

\begin{document}

\title{Bounds on the mass-to-radius ratio for non-compact field configurations}
\author{Shahar Hod}
\address{The Ruppin Academic Center, Emeq Hefer 40250, Israel}
\address{ }
\address{The Hadassah Institute, Jerusalem 91010, Israel}
\date{\today}

\begin{abstract}

\ \ \ It is well known that a spherically symmetric compact star
whose energy density decreases monotonically possesses an upper
bound on its mass-to-radius ratio, $2M/R \leq 8/9$. However, field
configurations typically will not be compact. Here we investigate
non-compact static configurations whose matter fields have a slow
global spatial decay, bounded by a power law behavior. These matter
distributions have no sharp boundaries. We derive an upper bound on
the fundamental ratio $\max_r\{2m(r)/r\}$ which is valid throughout
the bulk. In its simplest form, the bound implies that in any region
of spacetime in which the radial pressure increases, or
alternatively decreases not faster than some power law
$r^{-(\gamma+4)}$, one has $2m(r)/r \leq (2+2\gamma)/(3+2\gamma)$.
[For $\gamma \leq 0$ the bound degenerates to $2m(r)/r \leq 2/3$.]
In its general version, the bound is expressed in terms of two
physical parameters: the spatial decaying rate of the matter fields,
and the highest occurring ratio of the trace of the pressure tensor
to the local energy density.
\end{abstract}
\bigskip
\maketitle

\section{Introduction}

The spherically symmetric Schwarzschild black hole is characterized
by a mass-to-radius ratio of $2M/R=1$. However, this solution of the
Einstein equations has a central singularity, clothed by the
black-hole horizon at $R=2M$. It was Schwarzschild who already in
1916 asked the question: how large can $2M/R$ possibly be for a {\it
regular} field configuration?

This quantity is central for the determination of the spacetime
geometry of an Einstein-matter solution, and it is also a measure
for the strength of the gravitational interaction. In addition, a
bound on $2M/R$ has observational consequences, since it limits the
observable redshift of the compact object (star).

For a compact configuration with a constant energy density and
isotropic pressures Schwarzschild obtained the bound \cite{Sch}

\begin{equation}\label{Eq1}
2M/R=8/9\  .
\end{equation}
Beyond this value the star would not be able to support its own
gravitational field, and it would ultimately collapse to form a
black hole. The existence of such an upper bound is intriguing
because it occurs strictly before the appearance of an apparent
horizon at $2M/R=1$. Years later, Buchdahl \cite{Buch} proved that
the same bound holds for isotropic compact spheres in which the
energy density decreases monotonically. The bound (\ref{Eq1}) is
commonly known as the Buchdahl inequality, and is included in most
text books on general relativity, see e.g. \cite{Wal,Wein}.

Because of its astrophysical importance in determining the
gravitational redshift factor of a compact star \cite{NoteChris},
the Buchdahl bound has been investigated over the years by many
researches. It has been found that various bounds can be obtained on
the mass-to-radius ratio, $2M/R$, depending on the assumptions made
on the structure of the static configuration and its matter content,
see
\cite{Bondi64,Fra,Bond1,Baum,Mars,Schun,Guv,Annin,Bohmer,Haks,Hakq,Hak1,Giul,Kara}
and references therein.

Most former studies have considered compact bodies with sharp
boundaries, for which the energy density has support in some finite
region $[0,R]$, with the assumption that the radial pressure drops
to zero at the surface of the compact spherical object, $p(r=R)=0$.
However, field configurations may typically be non-compact,
characterized by energy densities and pressures that approach zero
only asymptotically.

In this work we would like to study this regime, of {\it
non}-compact slowly decaying matter distributions. Specifically, we
shall analyze the strength of gravity for non-compact configurations
in which the radial pressure increases locally, or alternatively
decreases not faster than some inverse power law $r^{-\alpha}$. In
these cases the matter fields approach zero only asymptotically, and
thus the corresponding configurations possess no sharp boundaries.

The rest of the paper is devoted to the investigation of how large
the fundamental mass-to-radius ratio, $2m(r)/r$, can be for these
slowly decaying field configurations. The paper is organized as
follows. In Sec. II we formulate the Einstein-matter equations in a
form which would be convenient for the analysis of the behavior of
the radial pressure. In Sec. III we obtain an upper bound on the
quantity $\max_r\{2m(r)/r\}$ for the canonical case, in which the
stress-energy tensor of the matter fields satisfies the commonly
used energy conditions. The results are extended in Sec. IV, where
we consider a generalized version of the energy condition. We
conclude in Sec. VI with a summary of the main results. We also
discuss the physical implications of the analytically derived bound.

\section{The Einstein-matter equations}

The metric of a static spherically symmetric spacetime takes the
following form in Schwarzschild coordinates \cite{Nun}

\begin{equation}\label{Eq2}
ds^2=-e^{-2\delta}\mu dt^2 +\mu^{-1}dr^2+r^2(d\theta^2 +\sin^2\theta
d\phi^2)\  ,
\end{equation}
where the metric functions $\delta$ and $\mu$ depend only on the
Schwarzschild radius $r$. Asymptotic flatness requires that as $r
\to \infty$,

\begin{equation}\label{Eq3}
\mu(r) \to 1\ \ \ and\ \ \ \ \delta(r) \to 0\  ,
\end{equation}
and a regular center requires \cite{Noter}

\begin{equation}\label{Eq4}
\mu(r) =1+O(r^2)\ \ \ and\ \ \ \ p(0)=p_T(0)\  .
\end{equation}

Taking $T^{t}_{t}=-\rho$, $T^{r}_{r}=p$, and
$T^{\theta}_{\theta}=T^{\phi}_{\phi}=p_T$, where $\rho$, $p$, and
$p_T$ are identified as the energy density, radial pressure, and
tangential pressure respectively \cite{Bond1}, the Einstein
equations $G^{\mu}_{\nu}=8\pi T^{\mu}_{\nu}$ reads

\begin{equation}\label{Eq5}
\mu'=-8\pi r\rho+(1-\mu)/r\  ,
\end{equation}

\begin{equation}\label{Eq6}
\delta'=-4\pi r(\rho +p)/\mu\  ,
\end{equation}
where the prime stands for differentiation with respect to $r$.

The mass $m(r)$ contained within a sphere of radius $r$ is given by

\begin{equation}\label{Eq7}
m(r)=\int_{0}^{r} 4\pi r'^{2} \rho(r')dr'\  .
\end{equation}
Taking cognizance of the Einstein equation (\ref{Eq5}) one finds
$\mu(r)=1-2m(r)/r$. (We use gravitational units in which $G=c=1$.)

The conservation equation, $T^{\mu}_{\nu ;\mu}=0$, has only one
nontrivial component \cite{Nun}

\begin{equation}\label{Eq8}
T^{\mu}_{r ;\mu}=0\  .
\end{equation}
Substituting Eqs. (\ref{Eq5}) and (\ref{Eq6}) in Eq. (\ref{Eq8}),
one finds for the pressure gradient

\begin{eqnarray}\label{Eq9}
p'(r)&=& {{1} \over {2\mu r}}\Big[(3\mu-1-8\pi
r^2p)(\rho+p)+2\mu T\nonumber \\
&& -8\mu p\Big]\ ,
\end{eqnarray}
where $T=-\rho+p+2p_T$ is the trace of the energy momentum tensor.
Below we shall analyze the behavior of the function $P(r;\gamma)
\equiv r^{\gamma+4} p(r)$, whose derivative is given by

\begin{eqnarray}\label{Eq10}
P'(r;\gamma)&=& {{r^{\gamma+3}} \over {2\mu}}\Big[(3\mu-1-8\pi
r^2p)(\rho+p)+2\mu T\nonumber \\
&& +2\mu\gamma p\Big]\ .
\end{eqnarray}

When analyzing the coupled Einstein-matter system, one usually
impose some energy conditions on the matter fields. Two commonly
used energy conditions are:

\begin{itemize}
\item{The weak energy condition (WEC). This means that the energy
density, $\rho$, is positive semidefinite and that it bounds the
pressures. In particular, one usually assumes $0\leq p \leq \rho$.}
\item{The trace of the pressure tensor plays a central role in determining
the spacetime geometry of the equilibrium configuration. It is
usually assumed to satisfy the relation $p+2p_T \leq \rho$ (see
\cite{Bond1} and references therein). This condition is likely to be
satisfied by most realistic matter models \cite{Bond1}, and in
particular it holds for Vlasov matter \cite{Vla1,Vla2}.}
\end{itemize}

\section{The canonical case}

In this section we obtain upper bounds on the fundamental ratio
${\max}_r\{2m(r)/r\}$ for general Einstein-matter models which
satisfy the canonical energy condition $p+2p_T \leq \rho$ ($T\leq
0$). Taking cognizance of Eq. (\ref{Eq9}) together with the energy
condition, one finds

\begin{equation}\label{Eq11}
P'(r;\gamma)\leq {{r^{\gamma+3}} \over
{2\mu}}\Big[(3\mu-1)(\rho+p)+2\mu\gamma p\Big]\ .
\end{equation}
Next, we may use the inequalities $\gamma p \leq \gamma (\rho+p)$
for $\gamma \geq 0$, and $\gamma p\leq 0$ for $\gamma \leq 0$ in Eq.
(\ref{Eq11}) and obtain

\begin{equation}\label{Eq12}
P'(r;\gamma)\leq {{r^{\gamma+3}} \over
{2\mu}}\Big[(3\mu-1)+2\mu\gamma\Theta(\gamma) \Big](\rho+p)\ ,
\end{equation}
where $\Theta(x)$ is the Heaviside step function.

We consider {\it non}-compact field configurations for which any
decline of the radial pressure, $p(r)$, is bounded by some power
law, $r^{-(\gamma+4)}$ \cite{Notedec}. This implies that the
combined pressure function, $P(r)$, is a monotonic increasing
function. Equation (\ref{Eq12}) therefore yields a lower bound on
$\mu(r)$,

\begin{equation}\label{Eq13}
{\min}_r\{\mu(r)\} \geq {1 \over {3+2\gamma\Theta(\gamma)}}\ ,
\end{equation}
which, in turn, implies an upper bound on the fundamental
mass-to-radius ratio:

\begin{equation}\label{Eq14}
{\max}_r \Big\{{{2m(r)} \over r}\Big\} \leq {{2+2\gamma} \over
{3+2\gamma}}\  ,
\end{equation}
for $\gamma \geq 0$, and

\begin{equation}\label{Eq15}
{\max}_r \Big\{{{2m(r)} \over r}\Big\} \leq {2 \over 3}\  ,
\end{equation}
for $\gamma \leq 0$.

We point out that for $\gamma \leq 3$ (that is, any decrease of the
radial pressure is bounded by $r^{-7}$), the newly derived upper
bound is stronger than the canonical Buchdahl inequality,
$\max_r\{2m(r)/r\} \leq 8/9$.

\section{Generalized energy condition}

Our results may be extended using a generalized energy condition of
the form (see \cite{Hak1} and references therein)

\begin{equation}\label{Eq16}
p+2p_T \leq \Omega\rho\  ,
\end{equation}
where $\Omega \geq 0$. This condition is very general. Indeed, a
realistic matter model which satisfies the dominant energy condition
(DEC) \cite{Haw} would satisfy this inequality with $\Omega=3$. We
shall now consider two distinct cases:

{\it Case I}:\ \ $\Omega \geq 1$\ .--- In this case one can
substitute the inequalities $T\leq (\Omega-1)\rho \leq
(\Omega-1)(\rho+p)$ and $\gamma p \leq \gamma(\rho+p)\Theta(\gamma)$
into Eq. (\ref{Eq10}) to obtain

\begin{equation}\label{Eq17}
\Big[(3\mu-1)+2\mu(\Omega-1)+2\mu\gamma\Theta(\gamma)\Big](\rho+p)\geq
0\  .
\end{equation}
This sets lower bounds on the function $\mu(r)$ which, in turn,
yield an upper bound on the maximal mass-to-radius ratio of the
configuration

\begin{equation}\label{Eq18}
{\max}_r \Big\{{{2m(r)} \over r}\Big\} \leq {{2\Omega+2\gamma} \over
{1+2\Omega+2\gamma}}\  ,
\end{equation}
for $\gamma \geq 0$, and

\begin{equation}\label{Eq19}
{\max}_r \Big\{{{2m(r)} \over r}\Big\} \leq {{2\Omega} \over
{1+2\Omega}}\  ,
\end{equation}
for $\gamma \leq 0$.

It is interesting to note that a self-gravitating configuration with
zero radial pressure \cite{Flor} which satisfies the energy
condition $2p_T=\Omega\rho$ may saturate the upper bound, Eq.
(\ref{Eq19}). Substituting these conditions into Eq. (\ref{Eq10}),
one indeed finds $\mu=(1+2\Omega)^{-1}$, which implies
$2m(r)/r=2\Omega/(1+2\Omega)$.

{\it Case II}:\ \ $0\leq\Omega\leq 1$\ .--- In this case one may use
the inequality $T\leq (\Omega-1)\rho \leq (\Omega-1)(\rho+p)/2$.
Substituting this into Eq. (\ref{Eq10}), one obtains

\begin{equation}\label{Eq20}
\Big[(3\mu-1)+\mu(\Omega-1)+2\mu\gamma\Theta(\gamma)\Big](\rho+p)\geq
0\  .
\end{equation}
From here one finds the upper bounds

\begin{equation}\label{Eq21}
{\max}_r \Big\{{{2m(r)} \over r}\Big\} \leq {{1+\Omega+2\gamma}
\over {2+\Omega+2\gamma}}\  ,
\end{equation}
for $\gamma \geq 0$, and

\begin{equation}\label{Eq22}
{\max}_r \Big\{{{2m(r)} \over r}\Big\} \leq {{1+\Omega} \over
{2+\Omega}}\  ,
\end{equation}
for $\gamma \leq 0$.

\section{Summary and Conclusions}

We have investigated the behavior of the fundamental mass-to-radius
ratio, $2m(r)/r$, for non-compact static configurations in which the
matter fields have a slow global spatial decay. Contrary to compact
bodies studied in the past, the non-compact configurations studied
here have no sharp boundaries, though their total mass is finite.

The physical importance of the non-compact field configurations
studied here lies in the fact that, in physical situations the
pressure function and its derivatives are expected to be continuous
analytic functions. On the other hand, a compact body has a
non-continuous pressure gradient at its surface. Namely, $p'(r)<0$
as $r\to R^{-}$, and $p'(r)=0$ as $r \to R^{+}$, where $R$ is the
radius of the compact body. It is therefore natural and highly
important to analyze the behavior of static field configurations for
which the matter content approaches zero only asymptotically. In
these cases, which we have studied here, the radial pressure and its
gradients are smooth analytic functions throughout the bulk.

We have shown that if the spatial decay of the radial pressure is
not faster than some inverse power law, $r^{-(\gamma+4)}$, then the
quantity $\max_r\{2m(r)/r\}$ is bounded from above by a simple
relation which depends on the power index $\gamma$ and on the
quantity $\Omega$ which bounds the ratio between the trace of the
pressure tensor to the local energy density. The newly derived upper
bound is summarized in Table \ref{Table1} for the various regimes of
the parameter space.

We point out that the upper bound on $\max_r\{2m(r)/r\}$ becomes
tighter monotonically as one decreases the value of the parameter
$\Omega$. This parameter controls the strength of the pressures that
prevent the static configuration from collapsing under its own
gravity. This result is in accord with common sense-- the smaller
are the pressures, the weaker must gravity be in order to allow the
existence of static regular configurations.

Likewise, one finds that the upper bound on $\max_r\{2m(r)/r\}$
becomes stronger as the value of $\gamma$ decreases (for $\gamma<0$
the bound becomes degenerate). The parameter $\gamma$ is central in
determining the pressure gradients. The smaller is the value of
$\gamma$, the smaller are the outward pressure gradients, and this
implies that the upper bound on the strength of gravity must also be
tighter in order to avoid complete gravitational collapse.

Finally, it is worth emphasizing that had we considered only local
behavior of the fields, our bound would still hold locally. That is,
in any region of spacetime in which the radial pressure increases,
or alternatively decreases not faster than some power law, the
fundamental mass-to-radius ratio, $2m(r)/r$, conforms to the bound
given by Table \ref{Table1}.

\begin{table}[htbp]
\centering
\begin{tabular}{|c|c|c|}
\hline
\ & $0\leq \Omega \leq 1$ & $1\leq \Omega$ \\
\hline
\ $\ \gamma \leq 0\ \ $\ \ & \ ${{1+\Omega} \over {2+\Omega}}$\ \ &\ ${{2\Omega} \over {1+2\Omega}}$ \\
\ & \ & \  \\
\ $\gamma \geq 0$\ \ & \ \  ${{1+\Omega+2\gamma} \over
{2+\Omega+2\gamma}}\ \ $  &\ \ ${{2\Omega+2\gamma}
\over {1+2\Omega+2\gamma}}\ \ $ \\
\hline
\end{tabular}
\caption{Upper bounds on the fundamental mass-to-radius ratio,
$\max_r\{2m(r)/r\}$, for the various regimes of the parameter space.
The radial pressure is assumed not to decrease faster than some
inverse power law, $r^{-(\gamma+4)}$, and the matter fields satisfy
the generalized energy condition $p+2p_T \leq \Omega\rho$.}
\label{Table1}
\end{table}

\bigskip
\noindent{\bf ACKNOWLEDGMENTS}
\bigskip

This research is supported by the Meltzer Science Foundation. I
thank Uri Keshet and Yael Oren for stimulating discussions.


\begin{thebibliography}{99}
\bibitem{Noter} From Eq. (\ref{Eq9}) below one finds $dp/dr=2(p_T-p)/r$ as $r\to 0$ (where $\mu \to 1$).
Regularity of the pressure function therefore requires that
$p(0)=p_T(0)$.
\bibitem{Notedec} We note that for the total configuration's mass to be finite, the
energy density (and as a consequence of the weak energy condition
also the radial pressure) should approach zero faster than $r^{-3}$
asymptotically, see Eq. (\ref{Eq7}). Thus, at the asymptotic region,
$r \to \infty$, one has $\gamma>-1$.
\end{thebibliography}
\end{document}